# Forces from Lipids and Ionic Diffusion: Probing lateral membrane effects by an optimized filter region of voltage dependent K$^+$ channels

Johann Summhammer [a*], Georg Sulyok [a], Gustav Bernroider [b] and Massimo Cocchi [c]

[a] Institue of Atomic and Subatomic Physics, Technische Universität Wien, Stadionallee 2, 1020 Vienna, Austria
[b] Departmanet of Biosciences, Universität Salzburg (retired), Salzburg, Austria;
[c] Department of Veterinary Medical Sciences, Università di Bologna, Bologna, Italy

\* Correspondence: johann.summhammer@tuwien.ac.at

*Abstract:* We investigate the possible influence of poly unsaturated fatty acids (PUFAs) in the lipid bilayer of cell membranes on the conduction properties of the selectivity filter of voltage gated ion channels. Any change of this conductivity can have major consequences on electrical signalling in brain and other cells. At the microscopic level a change in the concentration of PUFAs can cause a change in the pressure distribution within the lipid bilayer, and this in turn can lead to a change in the length of the selectivity filter. In order to estimate the consequences this might entail for the ion channel's conductivity we performed high resolution molecular dynamics (MD) simulations of the passage of K$^+$ ions and H$_2$O molecules through the selectivity filter of the KcsA potassium ion channel. Our results show that a change in length of the selectivity filter of as little as 4 %, independent of whether the filter is made longer or shorter, will reduce the K$^+$ ion current by around 50%. And further squeezing or stretching by about 10% can effectively stop the current. For instructive purposes and to facilitate further interpretation of the results, we give details of the force models employed in the MD simulations. We finally discuss the present results in the context of possible effects of membrane lipid compositions on Kv ion permeation.

1. Introduction

Functional aspects behind biological organizations strongly build on the electrostatic interaction between their constituent molecules. One prototypical example is provided by the structural and dynamical aspects of cell membranes. There the interplay between lipids and inserted membrane proteins can account for the regulation of many vital processes including electrical signalling in nerve cells. A long sequence of studies has demonstrated that changes in the local composition of cell membranes can strongly modulate the ion-conductive properties of inserted membrane proteins, including voltage-gated channels by 'lipo-electric interactions', see [1] for a recent review. However there are some natural restrictions in the experimental study of molecular interactions in the membrane. First, manipulations usually have to focus on isolated target molecules and this approach may not maintain their integrated membrane function that can be expected from their naturally highly disordered and dynamic environment in the living membrane. A helpful route to overcome this restriction has been made available by the addition of theoretical and computational tools to simulate the environment of highly resolved structures, setting out from '(low-temperature) snapshots' of single molecules and study molecular dynamics (MD) by modifications of their interactions within a highly mobile environment, e.g. [2, 3].

Initially this approach was somehow hindered by limited computational power. It turned out that the characteristic time scale in molecular dynamics in general, at equilibrated conditions, are very short in the range of fs ($10^{-15}$ sec), whereas experimentally observed events such as transmembrane currents establish at the microsecond to millisecond range ($10^{-6} - 10^{-3}$ sec). The possibility to bridge this large gap between very rare events and the typical time scale behind macroscopic ensemble averages of events has become attainable only more recently by increased and parallel computational resources that can span the time differences without loss of decisive resolution.

The present report demonstrates the implementation of such simulations and the application for the study of lateral membrane effects exerted by the lipid-environment on voltage gated ion channels. Among a wide range of possibilities that alter the lateral segregation of cell membrane molecules with possible consequences on the function of inserted proteins, the special role of poly-unsaturated fatty acids (PUFAs) has received particular attention [1,4,5]. PUFAs are found to directly regulate membrane channel function, with far reaching physiological and neuro-psychological consequences [8], although the precise mechanism is still somehow elusive [6]. The basic distinction that is frequently made under a situation as this is to distinguish between two types of effects, i.e. 'direct' or 'non-specific' effects [1]. Here, 'direct effects' are defined as lipo-electric interactions between charged groups on the lipid (e.g. the negative charge on lipid head group carboxyls) on one side and functional parts of the channel protein (e.g. the voltage sensor domain VSD of Kv channels) on the other [6,7]. 'Non-specific membrane effects' may operate very generally through the 'force from lipids' (ffl) principle [4], changing the conformation of channels through positive and negative pressure profiles. However, this distinction may eventually resolve as these effects can be narrowed down to the highest resolution of the target principle, the ionic-coordination and translocation through the atomic surrounding of the channel molecule. This is precisely the intention of the present approach.

We focus on examining the functionally decisive domain of $K^+$ channel proteins, the selectivity filter (SF), that has received an enormous attention ever since its crystallographic characterization around the year 1998 [9]. The SF is formed by a tetramer of subunits with a highly conserved sequence of amino acids and carbonyl groups oriented towards the narrow filter pore (*Figure 1*). This configuration allows for four transient binding sites for filter ions ('oxygen-cages') that are largely responsible for the domains preference for certain ions (e.g. $K^+$ over others e.g. $Na^+$). Selective and conductive properties are assumed to be optimized by the delicate geometries between interacting charges along the filter atoms. How this arrangement can engage in selective binding while maintaining high conduction rates has been disputed by a sequence of papers during the last years, e.g. [10,11]. However, a 'very high resolution' view on filter dynamics reveals a new aspect that has not been detected by employing potentials from mean force fields (PMF) methods and time-scales at the nano to micro-second range as previously. If ion translocation is simulated from first principle methods at atomistic resolution with the channel held in the open gate state as in the present approach, ion-transfer appears to be clearly confined to single 'bursts' with a characteristic time in the pico-second range. This picture meets both properties of the channel as it renders the filter-states 'non-conductive', but potentially selective over extended time intervals in the range of ns but also allows conduction rates to add up to about ~10 ions/ns and to the experimental diffusion limit of ~$10^8$ ions/sec [12]. Much of our focus herein will be on the role of the emerging optimized geometry underlying this conduction and the possible environmental effects that can distort the associated dynamics. In other words, we examine the extent to which the present atomistic MD model of the KcsA SF can respond to perturbations arising from its surrounding molecular forces exerted by lipid compositions.

First we provide a detailed description of the SF based on the prototypical KcsA motive and the implementation of the present MD approach with forces derived from first principles methods. Subsequently we discuss the emerging dynamics with special attention paid to an optimized structure and its sensitivity to linear distortions along its longitudinal transmembrane direction, as this distance is seen to shorten or extend depending on the gating state of the channel from atomic force microscopy [13]. Finally we discuss the results with relation to the 'force from lipids' and 'lipo-electric' concepts, particularly applied to the role of PUFA interactions.

## 2. Methods

**The Selectivity filter of the KcsA potassium channel**
The selectivity filter of the KcsA $K^+$-channel is formed by a five residue sequence, the amino acids Threonine (Thr75), Valine (Val76), Glycine (Gly77), Tyrosine (Tyr78), and Glycine (Gly79), commonly abbreviated as TVGYG, which is highly conserved from bacteria to human cells (Fig.1). It is located within the P loop of each of the four subunits of the tetrameric structure of the ion channel proteins. Under physiological $K^+$ concentrations, the selectivity filter residues arrange to form rings of carbonyl oxygen atoms directed roughly toward the center of the pore axis. Four adjoining sites can be defined and are usually designed as S1–S4. Dehydrated $K^+$ ions bind in these cage-like sites. Further binding sites for partially or fully hydrated $K^+$ ions have been also identified at the extracellular entrance (S0, $S_{ext}$) and towards the central cavity ($S_{cav}$) of the channel [14,15].

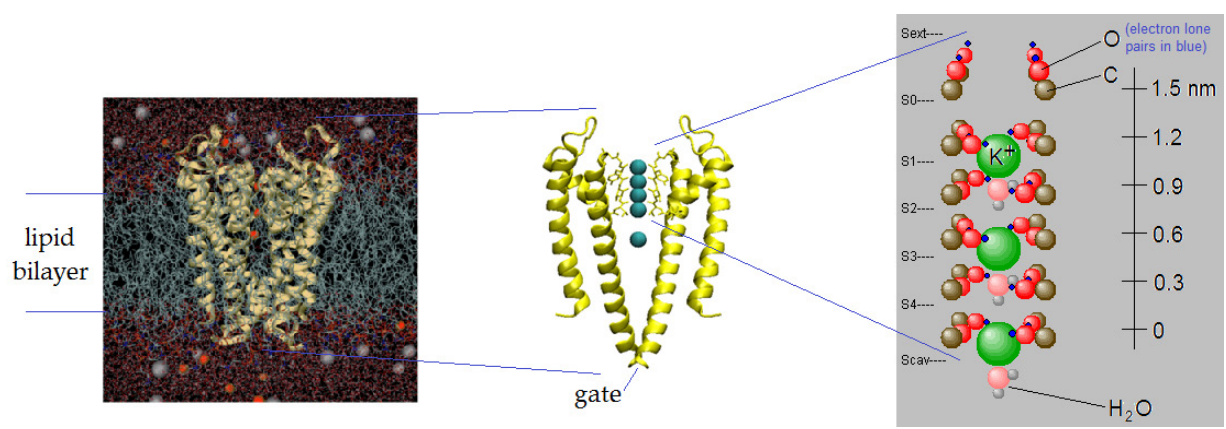

**Figure 1**: *Left:* A KcsA channel embedded in the cell membrane and surrounded by the lipid bilayer (adapted from [26]). Protein as yellow ribbons, lipid tails as chains, water as sticks and $K^+$ and Cl-ions as red and grey balls, respectively. Outside of the cell is on top, inside on bottom. $K^+$ ions can move from inside to outside through the narrow constriction posed by the selectivity filter, if the gate below the filter is open. *Middle:* Stripped view of the KcsA channel with the mechanical gate at the bottom and the selectivity filter occupied by $K^+$ ions (blue-green) in the upper part. Only two strands of the tetrameric protein structure are shown. *Right:* The selectivity filter occupied with a sequence of $K^+$ ions (green) and $H_2O$ molecules (pink and grey). Only the carbonyl groups forming the narrow pore of the selectivity filter are shown (carbons in brown, oxygens in red and their lone pairs electrons in blue). The preferred sites, at which the $K^+$ ions tend to be held by coordination to the oxygen atoms are indicated by S0 to S4. The sizes of the spheres representing the different atoms are as follows: Ion radius of $K^+$: 138 pm; Covalent radius of carbon (when in double bond with O): 64.1 pm; Covalent radius of oxygen (when in double bond with C): 58.9 pm; Center-center distance of C and O in carbonyl: 123 pm; Single bond covalent radius of O-atom within $H_2O$: 73 pm; Single bond radius of H-atom with $H_2O$: 38 pm; Center-center distance between O and H within $H_2O$: 95.84 pm.

**Modeling the selectivity filter**

Our model of the KcsA selectivity filter (SF) is shown on the right side of Figure 1. The SF is a nano pore of about 1.5 nm in length, which is made up of four strands of peptide chains. On each chain we find six essential carbonyl groups. We take the carbon atoms of these carbonyls (shown as brown spheres) to be at fixed positions. Each oxygen atom of a carbonyl (oxygens are shown as red spheres) has a constant distance to its carbon, but it can vibrate around its mean position in both angular directions. The carbon atoms have a positive partial charge and the oxygen atoms have a negative partial charge of equal magnitude (for details see the appendix). The charge of a carbon atom is centered, but the charge on the oxygen is dispersed over two points, in order to represent the lone pairs electrons of the oxygen, which play an important role in holding and transporting the $K^+$ ions and the water molecules which must diffuse through the SF. So one part of the charge of the oxygen is centered on the atom itself, but the lone pairs part is a point charge somewhat outside of the atom (shown as blue dots in Fig.1, right image). It is always located along the line of the C-O-axis, even when the O-atom swings around the C-atom.

The mobile species in the SF are the $K^+$ ions (green spheres) and the water molecules (pink oxygen atoms and attached grey hydrogen atoms. The $K^+$ ion carries one unit of positive charge. They can only move along the central axis of the SF. In the water molecules the oxygen atoms have a negative partial charge and the hydrogen atoms each have a positive partial charge of half the magnitude, so that the water molecule is neutral but has an electric dipole moment. The water molecules, too, can only move along the central axis of the SF. A water molecule can have two principal orientations: either the oxygen atom is on top, or a hydrogen atom is on top. In addition, a water molecule can have an arbitrary rotational angle around the central axis of the SF. Further details of the model, especially the forces acting between the various particles, are presented in the appendix.

**Motion of $K^+$ and $H_2O$ through the SF**

The calculation of the motion of $K^+$ and $H_2O$, as well as the vibrating motions of the oxygen atoms was done by methods of molecular dynamics based on the Verlet algorithm. The software was developed by us, in order to allow the incorporation of quantum mechanical motion of some of the particles later on, but quantum mechanical aspects are not of importance in the present context (Summhammer et al, to be published). The calculations we want to show here focused on the situation when the gate at the inner side of the membrane is open (Fig.1, middle picture) and ions and water should be able to flow easily from the inside of the cell to the outside. Therefore, there is a sufficient supply of $K^+$ ions and water molecules at the lower side of the SF. The $K^+$ in the cavity before the SF are usually in a hydrated state. For a $K^+$ to enter the SF, it must get rid of its shell of water molecules. During this process it is also possible for $H_2O$ molecules to enter the SF. Therefore, there is a pressure of $K^+$ as well as of $H_2O$ for entering the SF. These partial pressures are assumed to be equal and hence there is an equal chance for either a $K^+$ or an $H_2O$ to hop into site $S_{cav}$ as soon as this is sterically possible. The partial pressure of the $K^+$ ions is at the origin of the electric potential difference between the inside and the outside of the channel, because at the outside this pressure is kept at a smaller value. The motion of both species through the SF is mainly governed by the behavior of the $K^+$ ions, for which the SF is a series of potential barriers and valleys constituted by the electric charges. The $K^+$ have certain sites $S_{cav}$, S4,…,S0, where the lone pairs electrons of the oxygens of eight carbonyl groups coordinate around the ion, very similar to how water molecules would coordinate around a $K^+$ immersed in water. The $K^+$ can be attached to such a site for a relatively long time. However, in the model the whole structure of the SF is assumed to be in the thermal bath of the surrounding proteins, and therefore the O-atoms of the carbonyl groups exhibit random motion in addition to the motions which they undergo due to their interaction with the $K^+$, the water molecules, but also with other O-atoms and C-atoms. The temperature of this bath is kept at 310 K. Therefore, this thermal motion is also conveyed to the $K^+$ and to the $H_2O$, so that accidentally, they can overcome the barrier to the next site and hop into it. If at such a moment, all sites of the SF are occupied, this is only possible in a concerted action. If a site is free, then only a single $K^+$ or H2O from a neighboring site may move and hop into the site. Here, both directions of motion are possible, because the pure electric field due to the difference in $K^+$ concentration between inside and outside of the channel is too small to inhibit backwards motion. The motion of the $K^+$ ions through the SF is therefore a stop-and-go motion. In a typical KcsA channel in the open condition the ion current is on the order of several $10^8$ ions per second. On average, only every 2 to 5 nanoseconds a $K^+$ ion leaves the SF on the upper side, and since the occupancy at any given time is between 2 and 4 $K^+$ ions, an average ion spends between 6 and 15 nanoseconds in the SF. On the other hand it takes only around 1 picosecond for hopping the distance of around 0.3 nm from one site to the next. Theoretically, the SF could thus be traversed within 5 to 10 picoseconds, but in fact it takes about 1000 times longer. This is due to the strong coordination forces of carbonyl oxygens and their lone pairs electrons, which hold up a $K^+$ at every site within the SF. Figures 3(A) and 3(B) show two occupancy patterns of the SF which have an exceptionally long and a rather short hold up time of the $K^+$ current, respectively.

Water molecules as such could possibly traverse the SF on their own much faster, because their dipole forces do not pull them strongly to the carbonyls. But since they are mostly trapped between two $K^+$ ions, and passage through the SF is only possible in single files, their flow rate is essentially the same as that of the $K^+$ ions.

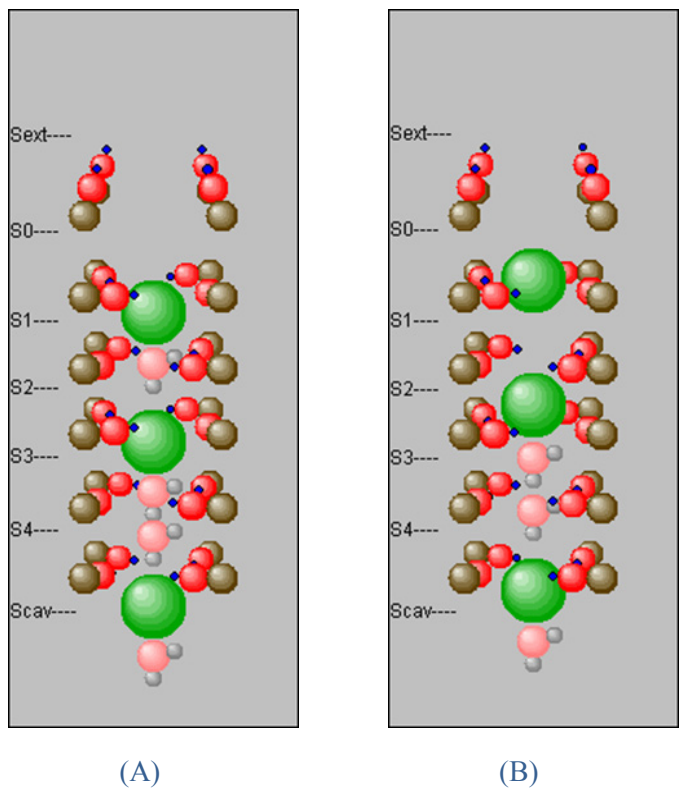

(A)　　　　　　　　(B)

**Figure 2:** (A) Example of a filter state (K1W2K3-WW4Kcav) which is characterized by extended stability with no ion emission over periods around 100 ns; (B) In contrast the occupancy pattern (K0K2WW4Kcav) is highly unstable and conductive, with inter-emission intervals around 1–2 ns.

**Simulating the effect of PUFA anomalies**

We have already seen in the introduction how an anomalous concentration of PUFAs can affect the gating behavior of potassium ion channels. Here we want to look at a possible influence on the selectivity filter of such channels, which is a much smaller structure. Just as in the larger structure of the ion channel as a whole, two effects can be envisioned. The changed concentration of PUFAs can either lead to the build-up of additional charges near the SF, or the changed local pressures can distort the geometry of the SF, and it is also conceivable that both effects act simultaneously. In the present paper we want to concentrate on the influence on the geometry, and in particular we want to look at stretching and squeezing of the SF along its central axis. In contrast to a radial squeezing or stretching, which would either lead to a quench of the ion current or to a strong increase when the nano pore becomes wider, it is not immediately clear what will happen when the sites S0 to S4, at which $K^+$ ions are held for long periods of time, are pulled farther apart or pressed closer together. Therefore we have made simulations in which the carbon atoms' relative distances parallel to the central axis of the SF were made larger or smaller. Since the oxygen atoms are tied to the carbons, this made the whole SF shorter or longer. Figure 3 gives a graphic presentation of the changes.

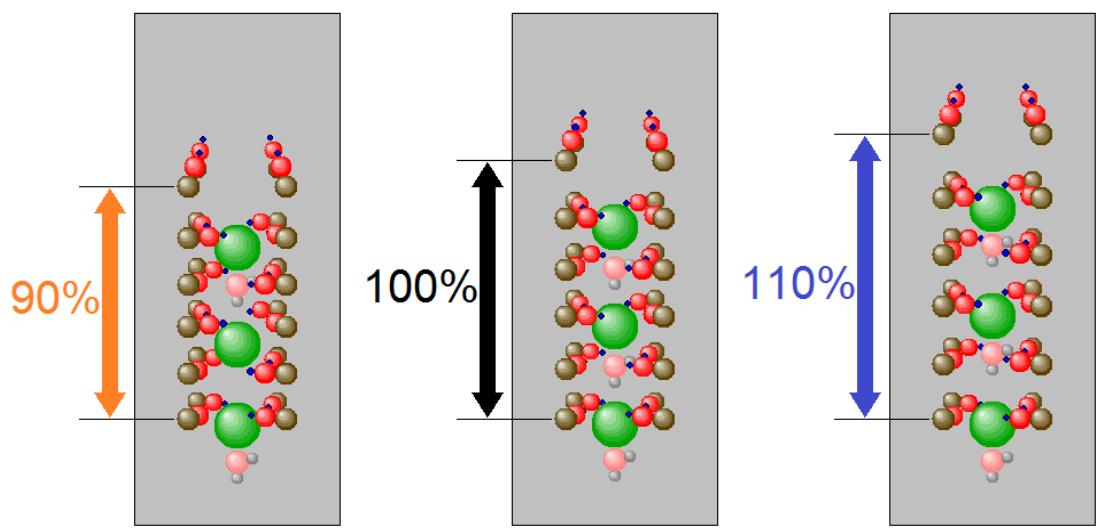

**Figure 3**: Changing the length of the selectivity filter. Simulations were made for different lengths, deviating in steps of 2% in the range of 90% to 110% of a 'standard length'.

At the beginning of a simulation of the current flow through the SF, the carbon atoms and with them the associated oxygen atoms were set to the positions appropriate for the chosen length for this particular simulation. Then $K^+$ ions and water molecules were placed into the SF as shown in the left configuration of Fig.4. Since this was certainly not a thermalized configuration, a thermalization period of 4 ns was started with a Verlet time step of 0.25 fs, so that a dynamical evolution over 4 million time steps could take place. During this period usually sufficiently many ions and water molecules had passed through the SF to have reached a thermal equilibrium of all the degrees of freedom of the involved particles. Often thermal equilibrium was reached within a much shorter time, as can be seen in the right configuration of Fig.4. After thermalization the actual simulation of interest was launched, which covered a period of 1 µs, again using a Verlet time step of 0.25 fs. The times of injection into and emission from the SF of $K^+$ ions during this evolution were recorded for statistical evaluation. For each of the investigated lengths of the SF between three and seven such simulations were made.

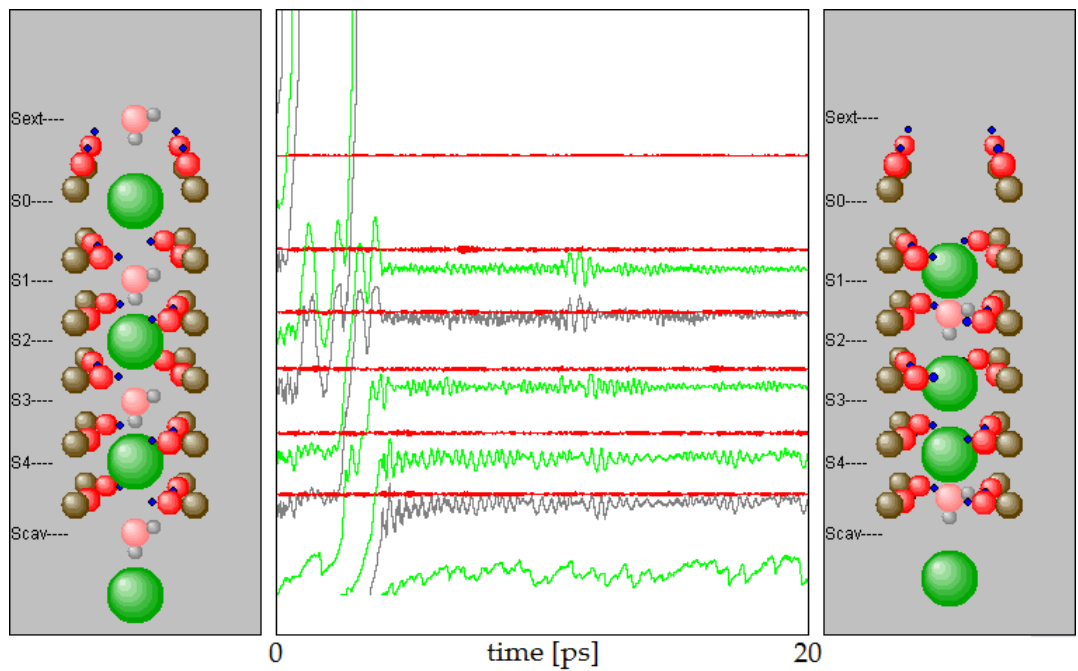

**Fig.4**: Initial configuration (left) and configuration after thermalization within 20 ps (right). The graph in the middle shows the z-component (upwards) of the momentary positions of $K^+$ ions (green), $H_2O$ molecules (grey) and O-atoms (red). Note that two $K^+$ and two $H_2O$ exit the SF on top within the first few ps. Within about the same time two $K^+$ enter below from the cavity region and an $H_2O$ and a $K^+$ get into the site $S_{cav}$ from the cavity, ready to enter the SF as soon as the particles at higher position move upwards. The vibrational amplitude of the O-atoms is relatively small due to the stiffness and corresponding high frequency of the C=O bond (wavenumber around 500 cm$^{-1}$).

## 3. Results

The present simulations provide insight to the question, whether stretching or squeezing of the selectivity filter can have an effect on $K^+$ ion conductance, given the atomic geometry of a standard open gate KcsA-SF model. Stretching or 'squeezing' are implemented as symmetrical variations of linear distances between backbone atoms along the longitudinal i.e. axial direction of filter atoms. Lateral forces influencing this geometry can be expected to arise mainly from the outer leaflet atoms of the surrounding lipid bilayer [1] (see discussion).

The fs-resolution MD model implemented at the atomic scale as reported here can detect the smallest changes in the vibrational modes of the interacting atoms. These modes come along with changes in the electronic distribution of neighboring atoms, e.g. with bending and stretching of carbonyl oxygen lone pairs as reported by density functional studies [16] and Brownian dynamics simulations [17] before. The interaction term between ions and electrons within the filter is therefore basically '*vibronic*', i.e. ionic motions change the electronic states of the molecule during their transitions. Details of this interaction will be discussed at another place. Here (e.g. **Figure 4**) we can observe oscillatory behavior of ions and intermittent water molecules in the range of THz (10$^{12}$ Hz). These oscillations 'couple' with almost zero-lag time correlations to their immediate neighbors in the filter under unperturbed conditions (e.g. during long periods without changes in the cavity location of ions, and ion exits from the filter). Such synchronizations become particularly stable in the K1,3 configuration (e.g. K1W2K3WW) where K denotes the $K^+$ ion, the number denoting the binding site location, and W represents water molecules, (see **Figure 2**). Relatively unstable configurations leading to ion exits from the channel mostly involve K2 and K4 occupancy states which is in accord with early studies suggesting a 'blocking-gate' in the SF with the K1,3 occupancy and all transits involving the K2,4 configuration [18].

**Figure 5** summarizes the results from simulations involving modifications of the longitudinal extension of the SF ranging from 90% to 110% in steps of 2%, including lengths of 95% and 105% of the standard length. For each extension three simulations were made, except for the standard length, which was used for seven simulation runs. There are at least two highly conspicuous findings: The first highly notable result is, that the $K^+$ ion current becomes maximal very close to the 'standard distance', this is at 101% of its standard length and that it drops off to about half of its normal value when the SF is extended or compressed by about 4% of its length. The second aspect is that the ion current exhibits a surprisingly high fluctuation under constant external conditions. For example at the standard extension of the SF currents can vary from only 28 ions per µs to as many as 387 ions per µs, with a mean current of 225 ions per µs. This is considerably higher than could be expected to arise from a Poisson type of ionic transduction building on a simple diffusion model. This aspect needs of course additional consideration given in the discussion section.

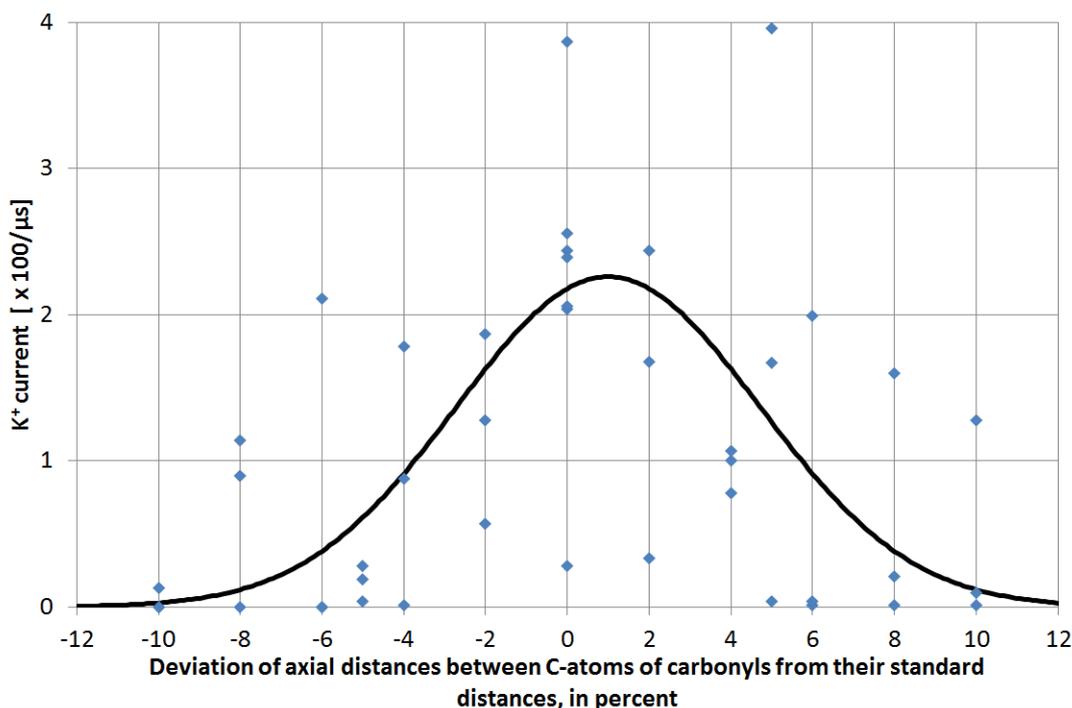

**Figure 5**: The $K^+$ ion current as a function of length of the selectivity filter. Each blue dot marks the result of a 1 µs molecular dynamics simulation of the selectivity filter resolved at 0.25 fs time steps. The black curve demonstrates a Gaussian fit to these data. (Fit function: $I(x) = I_0 e^{-(x-x_0)^2/w^2}$, with fit parameters $I_0 = 226$ ions/µs, $x_0 = 1$ percent, $w = 5.25$ percent.)

## 4. Discussion

We have demonstrated the structure and basic dynamical principles of the KcsA SF by an atomistic MD model employing a series of simulations, each of which extended over $4.10^9$ time steps of 0.25 fs and involved all fixed and mobile atoms with direct ion-interactions. In the course of manipulations of the geometrical properties underlying the control of conduction through filter atoms we have found an *optimized arrangement*. 'Optimized' in the present (mathematical) sense should be interpreted involving the interplay along three variables, the maximization of some function (a) provided by variables (b) in the context of a set of constraining principles (c). The function (a) is given by the conduction rate of the SF, the variables (b) are presented by the alignment of interaction distances in the SF and the constraining principles involve stability, selectivity and energetic principles of the SF. The results summarized in Figure 5 clearly identify the linear longitudinal extension along the alpha-helical backbone of four strands with five residues each lining the filter region, as the most sensitive variable controlling conduction. The 'standard' measure of this length maximizes conduction while maintaining stability and selectivity. Even a very small change in this length along both directions in the range of a few percent will significantly reduce conduction rates and changes around 10% of this length eventually stop conduction completely.

The high sensitivity of the filter length to changes in axial extension makes this measure particularly susceptible to perturbations from surrounding side chains, gating and lipid composition. The KcsA channel is known to shorten its longitudinal length upon opening the pore-gate of the protein. Beside the crystallographic picture the study of Sumino et al. confirms these changes by atomic force microscopy (AFM) of the naturally in-situ membrane-embedded channel [13]. These changes amount to about 20% with the 'short' version representing the open gate structure. To what extent these modifications upon gating will affect the geometry of the SF and the associated force field remains to be seen.

Another focus of attention must be laid on the electrostatic interaction of FAs, particularly PUFAs with the mobile parts of the protein such as the voltage-sensor domain (VSD). Reviewing this situation Elinder and Liin identify basically five sites of action for electro-static interactions on Kv channels [1]. At least two sites are close to the extracellular location of the filter and involve direct PUFA interactions, e.g. between carboxyl negative charges and positive gating charges on side chains surrounding the filter region, e.g. S4 [7]. The precise mechanism how these interactions translate into the complex channel dynamics still needs to be clarified. In particular, these studies depend on the 'temporal resolution' under which the events are studied. For example, the transition from a closed to an open-pore gate state of the channel involves several steps and each step can be characterized by a particular charge interaction and charge translocation associated with a rearrangement of atoms (e.g. during a helical screw motion in the VSD of Kv channels [19]). The characteristic life-time of this gating kinetics can be expected in the ms range. During this time frame the selectivity filter might change configuration between conductive and non-conductive states many times (about $10^3$ times as seen from a PMF model [18]). At this shorter µs time scale the filter states might still involve reorientations of peptide linkages distorting the geometry of the symmetrical coordination distances. However transitions shorter than µs remain unresolved in PMF studies. The present simulation goes much deeper resolving 1µs into four billion time steps. At this atomic scale we can identify periods in the range of 100 ns ($10^{-7}$s) where no ions are transmitted to the external site and there is no noticeable rearrangement of filter atoms during this period but a preferred K1,3 occupancy with intermittent W molecules. As mentioned in the introduction, the situation found here resembles 'dynamic gating' where permeation and selectivity result from the same process, a delicate situation that was anticipated early by the work of VanDongen 2004 [20].

Finally we want to make a note on the physiological significance behind the tight cooperation of an 'active membrane' regulating the interplay between lipids and ion channels as seen from the present work. The far reaching physiological effects of PUFAs together with their dietary role have received an enormous attention during the last years. In the context of neural function some PUFAs such as Linoleic Acid (LA) and alpha Linolenic acid (ALA) have been found to be of relevance for psychiatric conditions such as depression, as discussed with relation to evolutionary aspects by one of the co-authors, M. Cocchi [22]. However, a clearly defined site of action and the precise mechanism within the lateral heterogeneity of the membrane still have to be enlightened. In the traditional 'Fluid-Mosaic Model' of Singer and Nicholson [21] which has dominated the perspectives for almost half a century, these aspects are not well housed [5]. The noticeable low concentration of LA and ALA around 1-10μM in membranes has attracted different interpretations about a specific mechanism controlling channel function, particularly within the superfamily of Kv channels. Cocchi et al. provide a 'symmetry breaking' model in which low linoleic concentrations might operate at a boundary between 'phase-transitions' that eventually could affect larger membrane domains across cells and spread into the modular level of the entire brain [23]. Yet other evidence is supportive for direct effects between PUFAs and channel atoms because their PUFA effects can be 'washed out' within a few seconds and do not seem to have a significant fluidizing effect on the membrane [1]. However this situation will turn out eventually, the present work on a high resolution dynamic gating ability of the Kv filter domain has discovered a level of sensitivity regarding ion permeation that has not been seen previously. This level might still dissolve even more accurately by the combination of classical and quantum descriptions of ionic transition as suggested by the present authors in a number of contributions before [24,25]. Albeit still speculative, a connection between the level of conductive sensitivity of Kv channels as seen from the present results and the phase-transition hypothesis exerted by PUFAs as mentioned above, could possibly pave a path to the urgent question of channel cooperativity in the brain.

**References**

[1] Elinder F & Sara I. Liin, (2017) Actions and Mechanisms of Polyunsaturated Fatty Acids on Voltage-Gated Ion Channels, Frontiers in Physiology, 8.

[2] Roux B. & Schulten K. (2004) Computational Studies of Membrane Channels. Structure, 12, 13-43-1351.

[3] Miloshevsky G.V. & Jordan P.C. (2004) Permeation in ion channels: the interplay of structure and theory. Trends in Neuroscience, 27,6.

[4] Cordero-Morales JF & Vasquez V. (2018) How lipids contribute to ion channel function, a fat perspective on direct and indirect interactions. Current Opinion in Structural Biol, 51:92-98.

[5] Bagatolli L.A. & Mouritsen OG. (2013) Is the fluid mosaic (and the accompanying raft hypothesis) a suitable model to describe fundamental features of biological membranes? What may be missing? Frontiers in Plant Science, 4, 457.

[6] Yazdi,S. Stein M., Elinder F, Andersson M, Lindahl E. (2016) The Molecular Basis of Polyunsaturated Fatty Acid Interactions with the Shaker Voltage-Gated Potassium Channel. PLOS Computational Biology, 11.

[7] Börjesson, S. (2011) Polyunsaturated Fatty Acids Modifying Ion Channel Voltage Gating. (2011) Dissertation, Linköping University, Faculty of Health Sciences.

[8] Czysz A.H. & Rasenick M.M. (2013) G-Protein Signaling, Lipid Rafts and the Possible Sites of Action for the Antidepressant Effects of n-3 Polyunsaturated Fatter Acids. CNS Neurol Disord Drug Targets, 12(4), 466-473.

[9] Doyle DA, et al. (1998) The structure of the potassium channel: Molecular basis of $K^+$ conduction and selectivity. Science 280:69-77.

[10] Fowler PW et al. (2013) Energetics of Multi-ion Conduction Pathways in Potassium Ion Channels, JCTC, 9, 5176-5189.

[11] Köpfer DA et al. (2014) Ion permeation in $K^+$ channles occurs by direct Coulomb knock-on. Science, 346, 352-355.

[12] Roux B & Schulten K (2004) Computational Studies of Membrane Channels. Structure 12, 1343-1351.

[13] Sumino A, Sumikama T, Iwamoto ,M, Dewa1 T, & Oiki S. (2013) The Open Gate Structure of the Membrane-Embedded KcsA Potassium Channel viewed from the Cytoplasmic Side. Scientific Reports, 3, 1063.

[14] Mark S.P. Sansom, Indira H. Shrivastava, Joanne N. Bright, John Tate, Charlotte E. Capener, Philip C. Biggin, (2002) Potassium channels: structures, models, simulations, Biochimica et Biophysica Acta (BBA) - Biomembranes,Volume 156, 294-307

[15] Zhou, Y., Morais-Cabral, J., Kaufman, A. *et al.* (2001) Chemistry of ion coordination and hydration revealed by a $K^+$ channel–Fab complex at 2.0 Å resolution. *Nature* **414,** 43–48

[16] Guidoni, L. & Carloni, P. (2002) Potassium permeation through the KcsA channel: a density functional study. Biochimica et Biophysica Acta, 1563, 1-6.

[17] Chung SH, Corry B. (2007) Conduction properties of KcsA measured using Brownian Dynamics with flexible carbonyl groups in the selectivity filter. Biophysical Journal, 93, 44-53.

[18] Berneche S & Roux B. (2005) A gate in the selectivity filter of potassium channels. Structure 13, 591-600.

[19] Börjesson, S. I., and Elinder, F. (2011). An electrostatic potassium channel opener targeting the final voltage sensor transition. J. Gen. Physiol. 137, 563–577.

[20] VanDongen A.M.J. (2004). K channel gating by an affinity-switching selectivity filter. PNAS 101, 9, 3248–3252.

[21] Singer S.J. & Nicholson G.L. (1972) The fluid mosaic model of the structure of cell membranes. Science, 175, 720-731.

[22] Cocchi M, Tonello L. Gabrielli F. & Minuto Ch.(2014) Human and Animal Brain Phospholipids Fatty Acids, Evolution and Mood Disorders. J Phylogen Evolution Biol, 2:2

[23] Cocchi M, Minuto Ch, Tonello L, Gabrielli F, Bernroider G, Tuszynski J.A, Cappello F. and Rasenick M. (2017) Linolic acid: Is this the key that unlocks the quantum brain? Insights linking broken symmetries in molecular biology, mood disorders and personalistic emergentism. BMC Neurosci 18,38.

[24] Summhammer j, Sulyok G, Bernroider G. (2018) Quantum dynamics and non-local effects behind ion transition states during permeation in membrane channel proteins. Entropy, 20, 558.

[25] Bernroider G. & Summhammer J. (2012) Can quantum entanglement between ion-transition states effect action potential initiation? Cogn. Comput. 4, 29-37.

[26] Cournia Z, et al. (2015) Membrane Protein Structure, Function and Dynamics: A Perspective from Experiments and Theory. J. Membr. Biol. (Web June 11, 2015). **DOI:** 10.1007/s00232-015-9802-0

[27] Garofoli S & Jordan PC (2003) Modeling permeation energetics in the KcsA potassium channel. Biophys. J. 84(5) 2814–2830

**Appendix: Forces between the particles**

The 'particles' in the model of the selectivity filter are carbon atoms (C), oxygen atoms (O), potassium ions (K$^+$) and water molecules (H$_2$O). All of these particles carry full or partial charges, or have an electric dipole moment. Therefore there may be Coulomb attraction or repulsion between them. But if any two of them get too close and their electron shells begin to overlap, there will always be repulsion. In the following we will describe the most important mutual forces between the particles.

**Carbon atoms:** The C-atoms are fixed at the backbone protein strands and are rigidly immobile during a simulation. But as each C is a member of a carbonyl group (C-O), it carries a partial charge of +0.38 q$_0$ (q$_0$...electric charge unit), and therefore exerts a Coulomb force on other charged particles. It was not necessary to consider short range repulsive forces between the C-atoms and other particles, because no particle could get near enough to a C-atom.

**Oxygen atoms:** Each O-atom is a member of a carbonyl group and is attached to its C-atom with the known center-to-center distance of 0.123 nm. When at rest, each O-atom points in a certain direction from its C-atom. This direction is adjustable by the two angles of spherical polar coordinates. For the simulations these angles have been set to typical values found in the literature [27]. It should be mentioned that usually these angles are such that the C-O axis does not point straight to the axis of the selectivity filter, as one might expect if the O-atoms are to form cages for the K$^+$ ions. This can be seen in figures 1, 2 and 3. E.g., at the outermost positions the C-O axis points upwards, while for the inner C-O groups it is horizontal and points to either the left or the right of the axis of the selectivity filter. Each O-atom can perform rotational vibrations around its two rest angles. The frequency of these vibrations is assumed to be the same for the two angles, and also assumed to be the same for all carbonyl groups. It was set to 500 cm$^{-1}$ in accordance with values reported before [17]. These rotational vibrations also have a damping factor, which is supposed to model the dissipation of energy from the selectivity filter to the protein backbone. Such dissipation is necessary, because when a K$^+$ ion hops from one site to the next, they acquire kinetic energy when settling into the new site. But in general they can be caught by the new site only, if this energy is dissipated.
The O-atoms also connect all mobile species within the selectivity filter to a thermal bath. This is achieved by random accelerations or decelerations on the rotational vibrations of the O-atoms, such that the mean energy contained in these motions corresponded to k$_B$T/2 per degree of freedom. Here k$_B$ is Boltzmann's constant and the temperature T was set to 310 K.
In order that each carbonyl group is electrically neutral, each O-atom carries a charge of -0.38 q$_0$. This charge is modelled as two point charges, of which one resides in the center of the O-atom and the other is somewhat outside the covalent radius of the atom, but on the line defined by the C-O axis. It represents the effective charge of the electron lone pairs of the O-atom. In the simulations this charge had a distance of 1.4 covalent radii from the center of the O-atom and carried 70% of its partial charge. Thus a charge of -0.266 q$_0$ was in the lone pairs point and -0.114 q$_0$ in the center of the O-atom. The lone pairs charge point was assumed to be fixed on the line of the C-O axis, so it followed the rotational vibrations of the O-atom.

**K$^+$ ions:** Each K$^+$ carried an electric charge of +1 q$_0$. It interacted with other particles by the Coulomb force. But if another particle got closer to a K$^+$ than the sum of the covalent radius of that particle and the ion radius of K$^+$ (set at 0.131 nm), a constant repulsive force set in. The strength of this force depended on the kind of the other particle, which could be an electron lone pairs charge, an O-atom or a water molecule. Generally it was scaled such that a complete overlap of the center points of the two particles would consume energy of several hundred to a thousand $k_BT$. Therefore no substantial overlap between a $K^+$ and any other particle ever occurred during the simulations. The motion of the K+ ions was constrained along the central axis of the selectivity filter. (This constraint was necessary to keep computation time at a manageable level.)

**$H_2O$:** The water molecules are modelled as three point charges arranged in the usual triangular fashion. The point charge corresponding to the oxygen atom carries a partial charge of -0.7 $q_0$ and the two point charges corresponding to the hydrogen atoms each carried a charge of +0.35 $q_0$. The distance of the hydrogen charge points to the oxygen charge point was set at the known distance between the corresponding nuclei in the water molecule, 95.84 pm. The angle between the bonds of the hydrogens to the oxygen was set to 104.5°. In order that the water molecules would not penetrate any other particles, a water molecule was also endowed with a constant repulsive force, which set in whenever any other particle did get too close. The border surface within which this force started to act was assumed as spherical, in good agreement with the shape of the outer electron shell of the water molecule, and the center of this sphere was on the symmetry axis of the water molecule somewhat along the way between the oxygen and the hydrogens. Just like for the $K^+$ ions, the motion of the hydrogen molecules was constrained to the central axis of the selectivity filter. And in order to save further computational time, the orientation of a hydrogen molecule was limited. One of the O-H axes was always aligned with the central axis of the selectivity filter, with either the hydrogen or the oxygen atom pointing upwards. But the rotational orientation of the other hydrogen atom around this axis could be any value between 0 and 360°.

## Examples of forces between the particles

In the following we give some graphical examples of the forces between the particles. The action of the forces is explained in the figure captions.

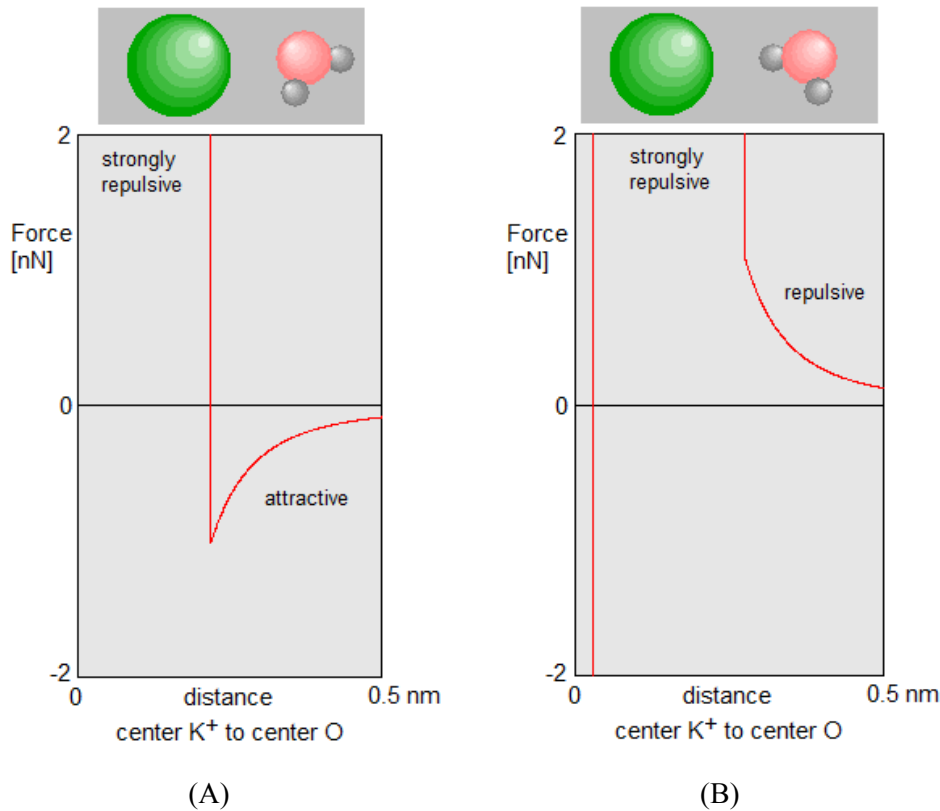

(A)                                   (B)

**Figure 6**: Force between a $K^+$ ion and a water molecule as a function of distance.
(A): The negative oxygen atom of the water molecule points towards the positive $K^+$ ion. As long as the distance between the center of $K^+$ and of the water molecule is larger than 0.248 nm (about 90% of the sum of the radius of $K^+$ and the effective radius of $H_2O$) the total coulomb force between the $K^+$ and the three charges of the water molecule is attractive with a maximum magnitude of almost $10^{-9}$ N. But within the repulsion zone an extremely strong repulsive force sets in. This configuration occurs when the water molecule is part of a hydration shell of the $K^+$ ion. When this configuration occurs within the selectivity filter, which is often the case when a water molecule from the inner cell cavity follows a $K^+$ ion into the selectivity filter, the water molecule gets attached to the $K^+$ and moves with it through the selectivity filter.
(B): One positive hydrogen atom of the water molecule points towards the positive K+ ion. Here the total Coulomb force between the $K^+$ and the water molecule is always repulsive, but at a distance of 0.5 nm it has fallen off to only about $2 \times 10^{-10}$ N. However, if the water molecule and the $K^+$ approach closer than the repulsion zone and the electron shells begin to overlap, the very strong repulsion force sets in.

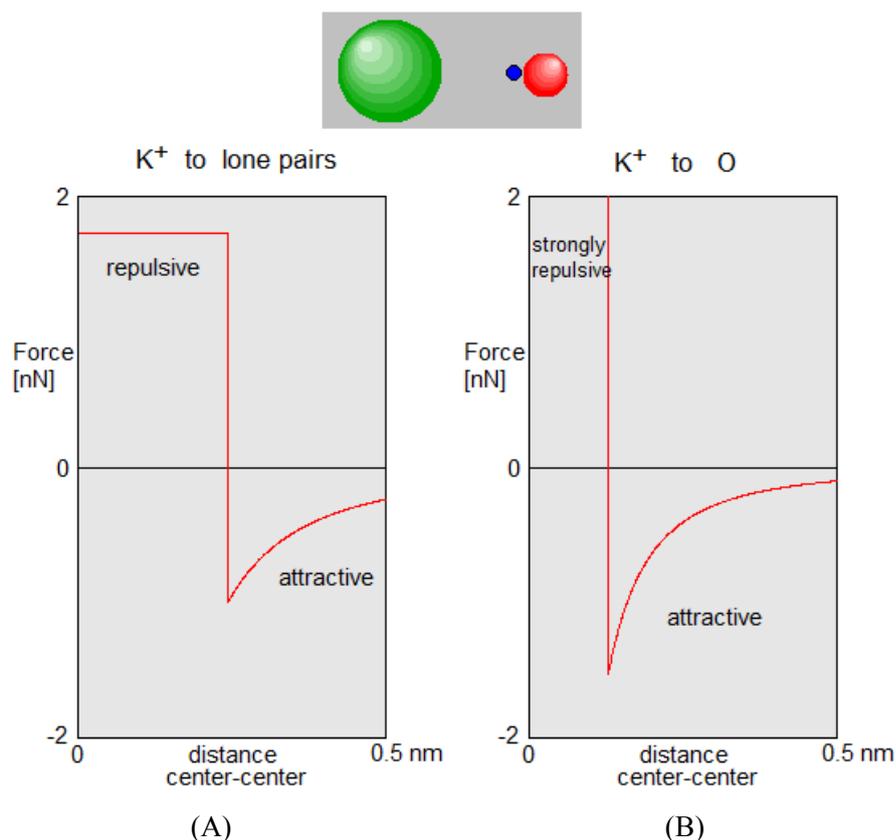

(A)                                  (B)

**Figure 7**: Force between a $K^+$ ion and the oxygen atom of a carbonyl group. As we model the oxygen atom as two point charges, where one also carries mass (the actual O-atom), and the other no mass but only a charge (the center point of the lone pairs charge distribution) one must consider separately the force between the lone pairs electrons of the oxygen atom and the $K^+$ (picture A) and the force between the actual oxygen atom and the $K^+$ (picture B).

(A): When the positive $K^+$ approaches the negative electron lone pairs point charge, there is a mutual attractive force due to the Coulomb interaction, which can reach almost $10^{-9}$ N. But at a center-center distance of the two charges of 0.238 nm the strong but constant repulsion of orbital overlap sets in, with a magnitude of about $1.7 \times 10^{-9}$ N, which corresponds to an energetic cost of full overlap of $K^+$ and the electron lone pairs charge density cloud of 100 $k_B T$ (T=310 K).

(B): When the positive $K^+$ approaches the negatively charged main body of the oxygen atom, the Coulomb force is attractive and reaches a value of about $1.5 \times 10^{-9}$ N. At center-center distances of $K^+$ and O of less than 0.124 nm the orbitals (other than the lone pairs electrons of the oxygen) begin to overlap and a very strong and constant repulsion of $1.72 \times 10^{-8}$ N sets in, which corresponds to an energetic cost of 500 $k_B T$ for full overlap of $K^+$ and O. During the simulations however, this repulsion never occurs, because the $K^+$ are constrained to the central axis of the selectivity filter and never get close enough to the oxygen atoms.

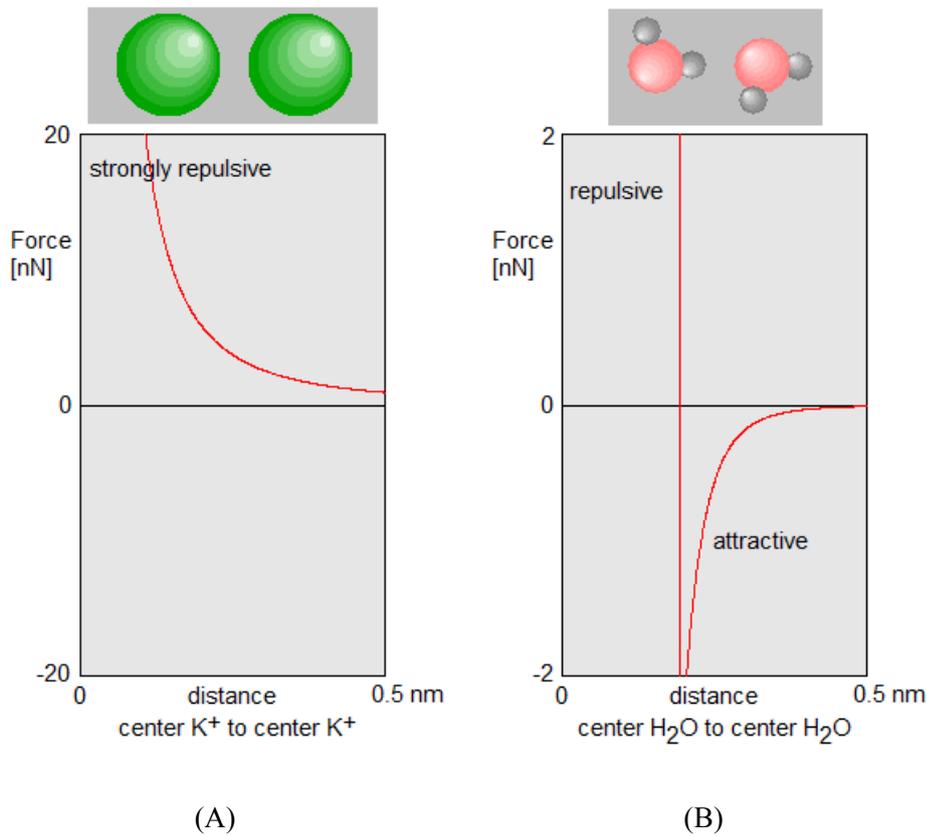

(A)                           (B)

**Figure 8:** (A) Force between two $K^+$ ions as a function of distance. This is a pure repulsive Coulomb force. Since each $K^+$ carries +1 q0, the repulsion is so strong that an energy of 5.22 eV, or almost 200 $k_B T$ at 310 K, would be required for the two charges to approach each other to a center-center distance of 0.276 nm, where theoretically additional repulsion due to overlap of electron shells should set in. Such a close approach practically never happens at this temperature, and therefore no additional repulsive force had to be taken into account.

(B): The force between two water molecules depends on their relative orientation and on their distance. Here we show the case where the negatively charged oxygen of one molecule points to one of the positively charged hydrogens of the other molecule. The resulting attractive force pulls the two molecules together until they enter the hydrogen bond. Any closer approach would cause a strong repulsive force.